\documentclass[aps,prl,floatfix,superscriptaddress,twocolumn]{revtex4-1}

\usepackage{amsmath}
\usepackage{bm}
\usepackage{graphicx}
\usepackage{color}

\newcommand{\ee}{\ifmmode (e,e^{\prime}) \else $(e,e^{\prime})$\fi}
\newcommand{\eep}{\ifmmode (e,e^{\prime}p) \else $(e,e^{\prime}p)$\fi}
\newcommand{\eepp}{\ifmmode (e,e^{\prime}pp) \else $(e,e^{\prime}pp)$\fi}
\newcommand{\eepn}{\ifmmode (e,e^{\prime}pn) \else $(e,e^{\prime}pn)$\fi}
\newcommand{\eepN}{\ifmmode (e,e^{\prime}pN) \else $(e,e^{\prime}pN)$\fi}
\newcommand{\pppn}{\ifmmode (p,2pn) \else $(p,2pn)$\fi}
\newcommand{\ppp}{\ifmmode (p,2p) \else $(p,2p)$\fi}
\newcommand{\heeepN}{\ifmmode {{}^4\textrm{He}(e,e^{\prime}pN)}
    \else ${}^4\textrm{He}(e,e^{\prime}pN)$\fi}
\newcommand{\gevcsq}{\ifmmode {(\textrm{GeV}/c)^2} \else $(\textrm{GeV}/c)^2$\fi}
\newcommand{\gevc}{\ifmmode {\textrm{GeV}/c} \else $\textrm{GeV}/c$\fi}
\newcommand{\mevc}{\ifmmode {\textrm{MeV}/c} \else $\textrm{MeV}/c$\fi}
\newcommand{\hefour}{\ifmmode {{}^4\textrm{He}} \else ${}^4\textrm{He}$\fi}
\newcommand{\ctwelve}{\ifmmode {{}^{12}\textrm{C}} \else ${}^{12}\textrm{C}$\fi}
\newcommand{\unsim}{\mathord{\sim}}

\begin{document}

\newcommand*{\JLAB}{Thomas Jefferson National Accelerator Facility, Newport News, VA 23606}
\newcommand*{\TLV}{Tel Aviv University, Tel Aviv 69978, Israel}
\newcommand*{\MIT}{Massachusetts Institute of Technology, Cambridge, MA 02139}
\newcommand*{\KENT}{Kent State University, Kent, OH 44242}
\newcommand*{\DOMINION}{Old Dominion University, Norfolk, VA 23529}
\newcommand*{\CALIF}{California State University, Los Angeles, Los Angeles, CA 90032}
\newcommand*{\HAMPTON}{Hampton University, Hampton, VA 23668}
\newcommand*{\PENNSYLVANIA}{Pennsylvania State University, State College, PA 16801}
\newcommand*{\Paris}{Institut de Physique Nucl\'{e}aire (UMR 8608), CNRS/IN2P3 - Universit\'e Paris-Sud, F-91406 Orsay Cedex, France}
\newcommand*{\Syracuse}{Syracuse University, Syracuse, NY 13244}
\newcommand*{\Kentucky}{University of Kentucky, Lexington, KY 40506}
\newcommand*{\William}{College of William and Mary, Williamsburg, VA 23187}
\newcommand*{\Virginia}{University of Virginia, Charlottesville, VA 22904}
\newcommand*{\Halifax}{Saint Mary's University, Halifax, Nova Scotia, Canada}
\newcommand*{\Glasgow}{University of Glasgow, Glasgow G12 8QQ, Scotland, United Kingdom}
\newcommand*{\Temple}{Temple University, Philadelphia, PA 19122}
\newcommand*{\Argonne}{Physics Division, Argonne National Laboratory, Argonne, IL 60439}
\newcommand*{\China}{China Institute of Atomic Energy, Beijing, China}
\newcommand*{\NRCN}{Nuclear Research Center Negev, Beer-Sheva, Israel}
\newcommand*{\Catania}{Universita di Catania, Catania, Italy}
\newcommand*{\Pittsburgh}{Carnegie Mellon University, Pittsburgh, PA 15213}
\newcommand*{\LongwoodUniv}{Longwood University, Farmville, VA 23909}
\newcommand*{\Florida}{Florida International University, Miami, FL 33199}
\newcommand*{\Tallahassee}{Florida State University, Tallahassee, FL 32306}
\newcommand*{\INFN}{INFN, Sezione Sanit\`{a} and Istituto Superiore di Sanit\`{a}, 00161 Rome, Italy}
\newcommand*{\INFNBari}{INFN, Sezione di Bari and University of Bari, I-70126 Bari, Italy}
\newcommand*{\Ohio}{Ohio University, Athens, OH 45701}
\newcommand*{\Tennessee}{University of Tennessee, Knoxville, TN 37996}
\newcommand*{\Kharkov}{Kharkov Institute of Physics and Technology, Kharkov 61108, Ukraine}
\newcommand*{\LOSALAMOS}{Los Alamos National Laboratory, Los Alamos, NM 87545}
\newcommand*{\Duke}{Duke University, Durham, NC 27708}
\newcommand*{\Texas}{University of Texas, Houston, TX 77030}
\newcommand*{\Seoul}{Seoul National University, Seoul, Korea}
\newcommand*{\Indiana}{Indiana University, Bloomington, IN 47405}
\newcommand*{\Hampshire}{University of New Hampshire, Durham, NH 03824}
\newcommand*{\Blacksburg}{Virginia Polytechnic Inst. and State Univ., Blacksburg, VA 24061}
\newcommand*{\France}{Universit\'e Blaise Pascal/IN2P3, F-63177 Aubi\`ere, France}
\newcommand*{\Mississippi}{Mississippi State University, Mississippi State, MS 39762}
\newcommand*{\Austin}{The University of Texas at Austin, Austin, Texas 78712}
\newcommand*{\Norfolk}{Norfolk State University, Norfolk, VA 23504}
\newcommand*{\Lanzhou}{Lanzhou University, Lanzhou, China}
\newcommand*{\Hebrew}{Racah Institute of Physics, Hebrew University of Jerusalem, Jerusalem, Israel}
\newcommand*{\Rutgers}{Rutgers, The State University of New Jersey, Piscataway, NJ 08855}
\newcommand*{\Yerevan}{Yerevan Physics Institute, Yerevan 375036, Armenia}
\newcommand*{\Ljubljana}{University of Ljubljana, Ljubljana, Slovenia}
\newcommand*{\Michigan}{Northern Michigan University, Marquette, MI 49855}
\newcommand*{\Hefei}{University of Science and Technology, Hefei, China}	
\newcommand*{\Jozef}{Jozef Stefan Institute, Ljubljana, Slovenia}
\newcommand*{\Ecole}{CEA Saclay, F-91191 Gif-sur-Yvette, France}
\newcommand*{\Massachusetts}{University of Massachusetts, Amherst, MA 01006}

\title{Probing repulsive core of the nucleon-nucleon interaction  \\ via the {\heeepN} triple-coincidence reaction.}

\author{I. Korover}
\affiliation{\TLV}
\author{N. Muangma}
\affiliation{\MIT}
\author{O. Hen}
\affiliation{\TLV}
\author{R. Shneor}
\affiliation{\TLV}
\author{V. Sulkosky}
\affiliation{\MIT}
\affiliation{\LongwoodUniv}
\author{A. Kelleher}
\affiliation{\MIT}
\author{S. Gilad}
\affiliation{\MIT}
\author{D.~W. Higinbotham}
\affiliation{\JLAB}
\author{E. Piasetzky}
\affiliation{\TLV}
\author{J.~W. Watson}
\affiliation{\KENT}
% ...

%\affiliation{Tel Aviv University\\ Israel}\\
\collaboration{The Jefferson Lab Hall A Collaboration}

\date{\today}
\label{dead}

\begin{abstract}

We studied simultaneously the ${}^4\textrm{He}\eep$, ${}^4\textrm{He}\eepp$, and
${}^4\textrm{He}\eepn$ reactions at\\ $Q^2=2~\gevcsq$ and $x_B>1$, for an
\eep\ missing-momentum range of 400 to 830 {\mevc}. The knocked-out proton
was detected in coincidence with a proton or neutron recoiling almost
back to back to the missing momentum, leaving the residual $A=2$ system at low excitation
energy. These data were used to
identify two-nucleon short-range correlated pairs and to deduce their
isospin structure as a function of missing momentum, in a region where
the nucleon-nucleon ($NN$) force is expected to change from predominantly
tensor to repulsive. The abundance of neutron-proton pairs is reduced as the nucleon momentum increases beyond
$\unsim500$ ${\mevc}$. The extracted fraction of proton-proton pairs is
small and almost independent of the missing momentum. Our data are compared with calculations of
two-nucleon momentum distributions in {\hefour} and discussed in the context of probing the elusive repulsive $NN$ force.
\end{abstract}


\maketitle

%\section{\label{sec:level1}First-level heading}
The stability of atomic nuclei is due to a delicate interplay between
the long-range attraction that binds nucleons and the short-range
repulsion that prevents the collapse of the system. In between, the
dominant scalar part of the nucleon-nucleon force almost vanishes and
the interaction is dominated by the tensor force, which depends on the
spin orientations and the relative orbital angular momentum of the
nucleons.

Recent high-momentum-transfer triple-coincidence \ctwelve\eepN\ and
\ctwelve\pppn\ measurements \cite{paper1,paper2,paper3,paper4} have
shown that nucleons in the nuclear ground state form pairs with large
relative momentum and small center-of-mass (CM) momentum, where large
and small are relative to the Fermi momentum of the nucleus. We refer to these pairs 
as short-range correlated (SRC)
pairs~\cite{paper5,paper6,paper20}. In the missing-momentum (the knocked-out proton initial 
momentum in the absence of final state interactions) range of $300-600~\mevc$,
these pairs were found to dominate the high-momentum tails of
the nuclear wave functions, with neutron-proton ($np$) pairs
nearly $20$ times more prevalent than proton-proton ($pp$) pairs, and by
inference neutron-neutron ($nn$) pairs. This is due to the strong
dominance of the $NN$ tensor interaction at the probed sub-fermi
distances~\cite{paper7,paper8,paper9}.

The association of the small \ctwelve\eepp\ / \ctwelve\eepn\ ratio, at
\eep\ missing momenta of $300-600~\mevc$, with dominance of the
$NN$ tensor force, leads naturally to the quest for increasing
missing momenta. This allows the search for pairs
at distances in which the nuclear force changes from being
predominantly tensor to the essentially unexplored repulsive
interaction. We report here on a simultaneous measurement of the
{\hefour}\eep, {\hefour}\eepp\ and {\hefour}\eepn\ reactions at
\eep\ missing momenta from 400 to 830 {\mevc}. The observed changes in
the isospin composition of the SRC pairs as a function of the missing
momentum are presented, discussed, and compared to calculations.

The experiment was performed in Hall A of Jefferson Laboratory (JLab)
using a 4 $\mu$A electron beam with an energy of 4.454 GeV incident on a 20-cm long
high pressure (13 atm, 20 K, 0.033  $\textrm{g}/\textrm{cm}^3$) {\hefour} gas 
target contained in a 8 cm diameter 20 cm-long aluminum cylinder.

The two Hall A high resolution spectrometers (HRS)~\cite{paper10}
were used to identify \hefour\eep\ events. Scattered electrons were
detected in the left HRS (L-HRS) at a central scattering angle of
$20.3^{\circ}$ and momentum of $3.602~\gevc$. This setup
corresponds to the quasi-free knockout of a single proton with
transferred three-momentum $\left| \vec{q} \right| \approx 1.64~\gevc$,
transferred energy $\omega\approx0.86$ GeV, the negative four-momentum transfer squared $Q^{2}\approx 2
{\gevcsq}$, and $x_B \equiv \frac{Q^2}{2m_p\omega} \approx 1.2$, where $m_p$ is
the proton mass. Knocked-out protons were detected using the
right HRS (R-HRS), which was set at three different central angles and
momenta: (33.5$^{\circ}$, $1.38~\gevc$), ($29^{\circ}$, $1.3~\gevc$),
and ($24.5^{\circ}$, $1.19~\gevc$).  These kinematical settings
correspond to \eep\ central missing momenta ($ \vec{p}_{\textrm{\textrm{miss}}} =
\vec{p}_{p} - \vec{q}$) values of $500~\mevc$, $625~\mevc$, and $750~\mevc$,
respectively, covering a missing-momentum range of $400-830~\mevc$
with overlap between the three different
settings.

The {\hefour}\eep\ events were selected by placing a $\pm 3\sigma$ cut around
the $\sigma = 0.6$ ns coincidence timing peak. The fraction of random events increased from $1 \%$ at the lowest missing 
momentum measurement to $9 \%$ at the highest. The other cuts on the \eep\ data 
were the nominal HRS phase-space cuts on
momentum ($|\Delta p/p| \le 0.045$) and angles ($\pm 60$ mrad vertical,
$\pm 30$ mrad horizontal). To reduce the random-coincidence background, a
cut on the target-reconstructed vertex ensured that both the electron and the proton emerged from the
same place within $\pm 3$ cm.
The $\Delta(1232)$
excitation was excluded by a cut on the quasi elastic \eep\ peak, as in Ref.~\cite{paper13}.

For highly correlated pairs, the missing momentum of the
$A\eep$ reaction is expected to be balanced almost entirely by a single
recoiling nucleon. A large acceptance spectrometer (BigBite) followed
by a neutron detector (HAND) with a matching solid angle was used to
detect correlated recoiling protons or neutrons. The experiment triggered on $e-p$ coincidences between the HRS spectrometers, 
with the BigBite and HAND detectors read out for every trigger.

The recoiling protons were detected by the BigBite spectrometer~\cite{paper11} 
centered at an angle of $97^{\circ}$, for the $500$ and $625~\mevc$ measurements, and $92^{\circ}$ for the $750~\mevc$
measurement. The angle between $\vec q$ and the recoil nucleon was $40^{\circ} - 50^{\circ}$. The angular acceptance was about $96$ msr and the
detected momenta accepted ranged from $0.25~\gevc$ to $0.90~\gevc$. The momentum
resolution of BigBite, determined from elastic electron-proton
scattering, was $\Delta p/p = 1.5\%$. The overall proton
detection efficiency was
$73 \pm 1\%$.

HAND consists of several elements: a 2.4-cm thick 
lead shield (to block low-energy photons and most of the charged particles coming 
from the target), followed by 64 2-cm thick scintillators (to identify and veto charged particles),
 and 112 plastic scintillator bars arranged in six 10-cm thick layers 
covering an area of 1 $\times$ 3 m$^2$ (to detect the neutrons). HAND was placed six meters from the
target, just behind BigBite, covering a similar solid angle as
BigBite.

The pattern of hits in sequential layers of HAND was used to identify
neutrons~\cite{paper14}. A time resolution of
$1.5$~ns allowed determination of the neutron momentum with an
accuracy that varied from $2.5\%$ (at $400~\mevc$) to $5\%$ (at $830~\mevc$).
The detection efficiency was $40 \pm 1.4\%$ for
$400-830~\mevc$ neutrons. This determination is based on the efficiency
measured up to $450~\mevc$ using the $d\eepn$ reaction, and extrapolated
using a simulation that reproduces well the measured efficiency at
lower momenta~\cite{paper15}.

\begin{figure}
\includegraphics[width=0.45\textwidth,height=2.8in]{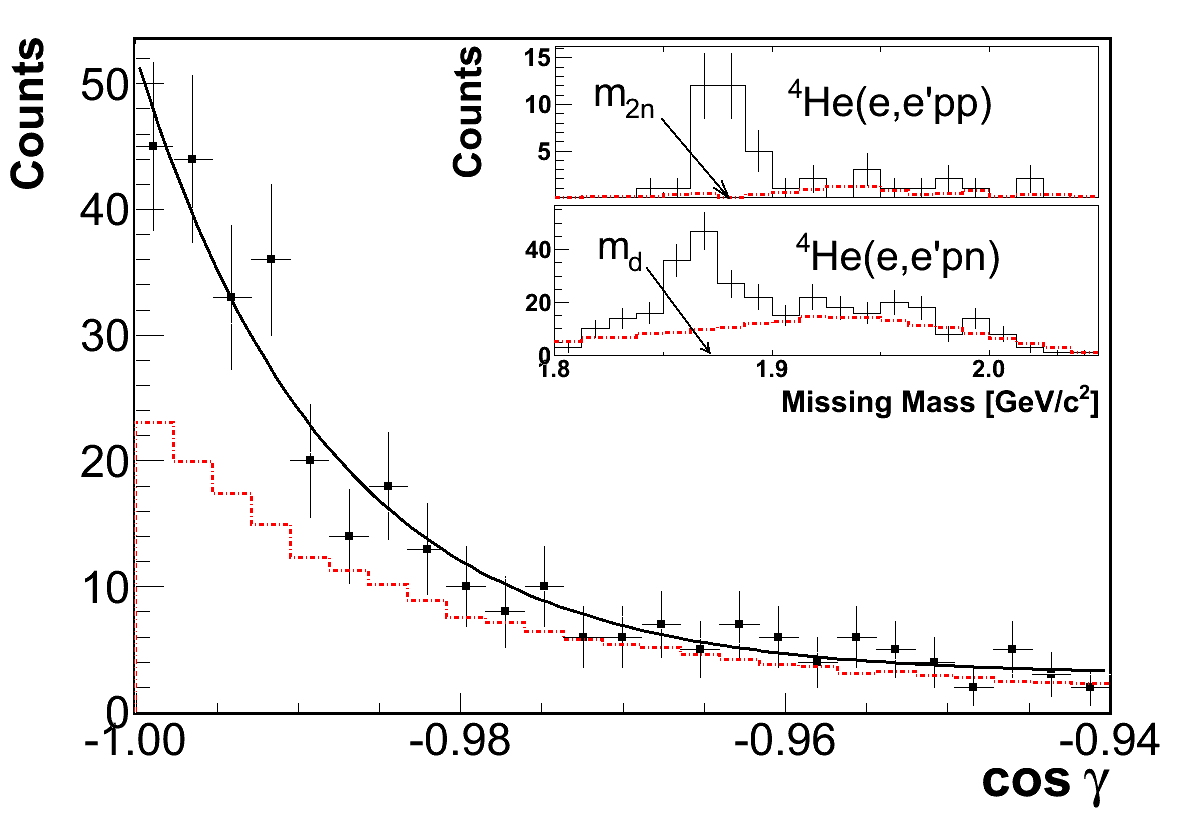}
\caption{The distribution of the cosine of the opening angle $\gamma$
  between the $\vec{p}_{\textrm{miss}}$ and $\vec{p}_{\textrm{recoil}}$ for the \hefour\eepn reaction
  ($p_{\textrm{miss}}=625$ and 750~{\mevc} kinematics combined). The solid curve is a simulation of scattering off a moving
  pair with a CM momentum having a width of 100 {\mevc}. The inserts show the missing-mass distributions. 
In both the main figure and the inserts the data is shown with no random background subtraction.
 The random background is shown as dashed dotted (red online) curves.}
\label{fig2} 
\end{figure}

The picture of 2N-SRC pair breakup with the other
two nucleons in \hefour\ being essentially spectators is supported by Fig.~\ref{fig2}.
The figure shows the distribution of the cosine of the
angle between the missing momentum and the recoiling neutrons ($\gamma$).  We
also show the angular correlation for the random background (dashed-dotted) as defined
by a time window off the coincidence peak.
While the placement of the neutron detector opposite to the nominal
missing momentum defined by the central rays of the high
resolution spectrometers leads to a geometrical
angular correlation in the random background, the real triple coincidence
events show a clear back-to-back peak above this background.
The curve
is a result of a simulation of the scattering of a moving pair as discussed
below. Similar back-to-back correlations were observed for the
recoiling protons. The inserts to Fig.~\ref{fig2} 
shows the missing-mass for the \hefour\eepp\ and \hefour\eepn\ reactions 
corresponding to a two-nucleon residual system with a low excitation energy.

Software cuts were applied to both BigBite and HAND that limited their
acceptances to  $\pm 14^\circ$ in the vertical direction,
$\pm 4^\circ$ in the horizontal direction, and $300-900~\mevc$ in momentum.
 A simulation based on the measurements was used to
correct the yield of the \hefour\eepN\ events for
the finite acceptances of the recoiling protons and neutrons in
Bigbite and HAND.  Following Ref.~\cite{paper1}, the simulations
assumed that an electron scatters off a moving SRC pair with a
center-of-mass (CM) momentum relative to the $A-2$ spectator system described by a
Gaussian distribution as in Ref.~\cite{paper16}.  We assumed an isotropic
$3$-dimensional motion of the pair and varied the width of the
Gaussian equally in each direction until the best agreement
with the data was obtained. The
nine measured distributions (three components in each of the three
kinematic settings for $np$ pairs) yield, within the uncertainties,
the same width with a weighted average of $100 \pm 20~\mevc$. This is in good agreement with the CM momentum distribution calculated in Ref.~\cite{paper9}.
Fig.~\ref{fig2} compares the simulated and measured
distributions of the opening angle between the knocked-out and
recoiling nucleons.  The fraction of events detected within the finite
acceptance was used to correct the measured yield. The uncertainty in
this correction was typically $15\%$, which dominates the systematic
uncertainties of the \hefour\eepN\ yield.

\begin{figure}
\includegraphics[width=0.45\textwidth,height=3.4in]{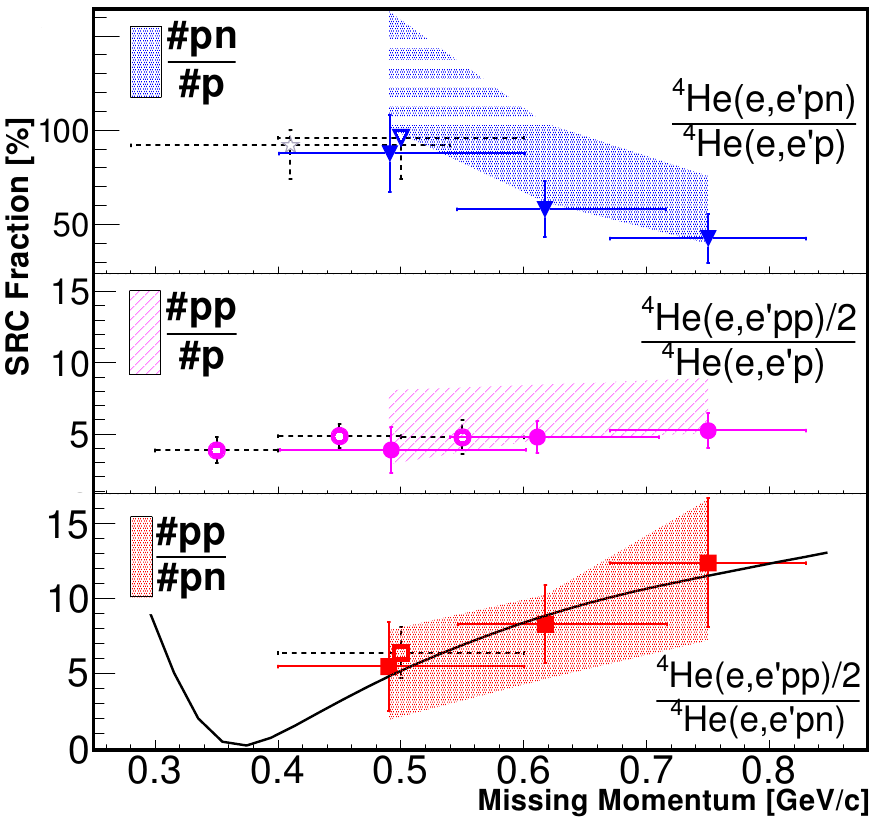}
\caption{ Lower panel: The measured ratios \hefour\eepp/
  \hefour\eepn\, shown as solid symbols, as a
  function of the $\hefour\eep$ missing momentum. Each point is the result of a different setting of the detectors.
  The bands represent the data corrected for FSI to obtain the pair
  ratios, see text for details. Also shown are calculations using the
  momentum distribution of Ref.~\cite{paper9} for pairs with weighted-average CM momentum assuming
  arbitrary angles between the CM and the relative momenta in the pair
  (solid black line).
  The middle panel shows the measured {\hefour}\eepp /
  {\hefour}\eep\ and extracted $\#pp/\#p$ ratios. The upper panel shows the measured {\hefour}\eepn
  /{\hefour}\eep\ and extracted $\#pn/\#p$ ratios. The unphysical region above $100\%$ obtained due to 
systematic uncertainties and statistical fluctuations is marked by white strips. Ratios for {\ctwelve} are shown as empty symbols
  with dashed bars. The empty star in the upper panel is the BNL result \cite{paper4} for \ctwelve\pppn/\ctwelve\ppp. 
 }
\label{fig:wide}
\end{figure}

The measured $\frac{\hefour\eepN}{\hefour\eep}$ ratios are given by the number of events in the
background-subtracted triple-coincidence TOF peak corrected for the finite acceptance and detection
efficiency of the recoiling nucleons, divided by the number of random-subtracted (double-coincidence) 
\hefour\eep\ events. These ratios, as a function of $p_{\textrm{miss}}$ in the
\hefour\eep\ reaction, are displayed as full symbols in the two
upper panels of Fig.~\ref{fig:wide}. Because the electron can scatter from either proton of a $pp$ 
pair (but only from the single proton of an $np$ pair), we divided the \hefour\eepp\ yield by two. Also displayed in Fig.~\ref{fig:wide},
 as empty symbols with dashed bars, are similar ratios for \ctwelve~obtained from previous electron scattering
~\cite{paper1,paper2} and proton scattering~\cite{paper4} measurements. In comparing the \ctwelve\ and \hefour\ data, it is noted that the measured ratios are about equal and very different from the ratios of naive pair counting in these nuclei.
The horizontal bars show the overlapping momentum acceptance ranges of the various kinematic settings. 
The vertical bars are the uncertainties,
which are predominantly statistical.

Because we obtained the \hefour\eepp\ and {\hefour}\eepn\ data
simultaneously and with the same solid angles and momentum
acceptances, we could also directly determine the ratio of
\hefour\eepp\ to {\hefour\eepn}. In this ratio, many of the
systematic factors needed to compare the triple-coincidence yields
cancel out, and we need to correct only for the detector
efficiencies. This ratio as a function of the missing momentum is
displayed in the lower panel of Fig.~\ref{fig:wide} together with the previously
measured ratio for \ctwelve~\cite{paper2}.

%\begin{figure*}[ht!]

To extract from the measured cross section ratios the underlying pair 
ratios, corrections for final-state interactions (FSI) were calculated using the Glauber
approximation~\cite{paper17}. The Glauber corrections ($T_{L}=0.75$ and
$T_{R}=0.66-0.73$), with $T_{L}$ and $T_{R}$ the leading and recoil transparencies, 
were calculated by the Ghent group ~\cite{paper17}. We assumed
the uncertainties to be $\pm 20\%$ of these values. The single charge exchange (SCX) probability ($P_{\textrm{SCX}}$) was
assumed to be $1.5 \pm 1.5\%$ based on the SCX total cross section of
$1.1 \pm 0.2$ mb ~\cite{paper18}. The pair fraction extracted from the measured ratios with the FSI calculated corrections are
shown in Fig.~\ref{fig:wide} as bands (see appendix for details). The statistical and systematic uncertainties were treated as independent
 and combined by simulation to create the width of the
one standard deviation bands shown in Fig.~\ref{fig:wide}. The systematic uncertainties in the correction factor 
($ 15\% $ due to finite detector acceptance, $ \sim 20\%$ due to FSI) 
and the statistical fluctuation can explain the extention of the band beyond $100\%$.

The correction to the ratios
due to attenuation of the leading-proton is small. The attenuation of the
recoiling nucleon decreases the measured triple/double coincidence
ratios. Because the measured \hefour\eepn\ rate is about an order of
magnitude larger than the \hefour\eepp\ rate,
\hefour\eepn\ reactions followed by a single charge exchange (and
hence detected as {\hefour\eepp}) increase the
\hefour\eepp/\hefour\eepn\ and the
\hefour\eepp/\hefour\eep\ measured ratios.

The two-nucleon momentum distributions were calculated for the ground
states of {\hefour} using variational Monte-Carlo wave functions derived
from a realistic Hamiltonian with Argonne V18 and Urbana X
potentials~\cite{paper9}. 
The solid (black) curve in Fig.~\ref{fig:wide} was obtained using the calculations ~\cite{paper9} weighted
average over arbitrary angles between $\vec K_{\textrm{rel}}$ and $\vec K_{\text{CM}}$,
the CM momentum of the pair. The calculation with $K_{\text{CM}}=0$, which agrees quantitatively with the Perugia 
group calculation~\cite{paper19}, differs little from the average shown in the figure.
To compare the calculations to the data in Fig.~\ref{fig:wide} we assumed that the virtual photon 
hits the leading proton and $p_{\textrm{miss}}=K_{\textrm{rel}}$ (Plane Wave Impulse Approximation).

The measurements reported here were motivated by the attempt to study the isospin 
decomposition of 2N-SRC as a proxy to a transition from primarily tensor 
to the short range repulsive, presumably scalar, nucleon-nucleon force. In the ground
state of \hefour\ \cite{paper9}, the number of $pp$-SRC pairs is much smaller
than $np$-SRC pairs for values of the relative nucleon momentum
$K_{\textrm{rel}} \approx400~\mevc$. This is because the correlations induced by the
tensor force are strongly suppressed for $pp$ pairs which are
predominantly in ${}^1S_0$ state \cite{paper7,paper8,paper9, paper19}.
As the relative momenta increase, the tensor force becomes less dominant,
the role played by the short-range
repulsive force increases and with it the ratio of $pp/np$ pairs.  In our measurement,
as the missing momenta is increased beyond 500~{\mevc}, the triple coincidence
\hefour\eepp/\hefour\eepn\ 
ratio increases, in good agreement with the prediction based on the ratio of
$pp$-SRC/$np$-SRC pairs in the \hefour\ ground state \cite{paper9}.

The measured triple/double coincidence ratios
shed further light on the dynamics. The measured \hefour\eepp/\hefour\eep\ ratio
reflects a small
contribution from $pp$-SRC pairs. These pairs are likely dominated by a scalar repulsive
short-range force which is relatively constant over the reported momentum range.

The \hefour\eepn/\hefour\eep\ ratio clearly shows that the
reduction in the $np$/$pp$ ratio with increasing $p_{\textrm{miss}}$
is due to a drop in $np$-SRC pairs with increasing $K_{\textrm{rel}}$.
While $np$-pairs 
still dominate two nucleon SRC, even at missing momentum of 800 \mevc\,
the total fraction of the \eep\ cross section associated with scattering from 
2N-SRC pairs drops with increasing missing momentum.  This is likely due to
an increase of more complex mechanisms,
such as stronger FSI and the onset of SRC involving more than
two nucleons~\cite{paper5}. A definitive understanding of the relative importance
of these effects requires exclusive measurements at large missing momentum on heavier nuclei,
and a more detailed theoretical study.

To summarize, the short range part of the $NN$ force is 
empirically known to be repulsive, it is essential to describe $NN$ scattering and
stability of nuclei, but it is difficult to explore and poorly known both theoretically
and experimentally. The measurements reported here probe a transition from 
an attractive to a repulsive $NN$ force. The data set, interpreted as changes in 
the isospin decomposition of the SRC pairs, is consistent with a reduced 
contribution from a tensor component and a constant contribution from a scalar
component of the $NN$ force over the probed missing momentum range. It
confirms the phenomenological description of the $NN$ force in this range.

One should  question to what level the naive interpretation of the data in terms
of the ground state nuclear properties is appropriate. Comprehensive calculations,
which take into account the full reaction mechanism in a relativistic treatment,
as well as additional data with better statistics will allow a more detailed determination
of the role played by the elusive repulsive short-range nucleon-nucleon interaction.

We acknowledge the contribution of the Hall A
collaboration and technical staff. We thank C. Colle, W. Cosyn and J. Ryckebusch
for the Glauber Calculations. We also want to thank R.B. Wiringa, R. Schiavilla, S. Steven,
and J. Carlson for the calculations presented in Ref.~\cite{paper9} that
were provided specifically for this paper. Useful discussions with J. Alster,
C. Ciofi degli Atti, W. Cosyn, A. Gal, L. Frankfurt, J. Ryckebusch,
M. Strikman, and M. Sargsian, are gratefully acknowledged. This work
was supported by the Israel Science Foundation, the U.S. National Science
Foundation, the U.S. Department of Energy grants DE-AC02-06CH11357,
DE-FG02-94ER40818, and 
U.S. DOE Contract DE-AC05-060R23177
under which Jefferson Science Associates
operates the Thomas Jefferson National
Accelerator Facility.

\appendix*
\section{Appendix}

To extract the SRC pair ratios ($\#pp/\#np$, $\#pp/\#p$, and
  $\#np/\#p$) from the measured cross-section ratios
  ($R=\frac{\hefour\eepp}{\hefour\eepn}$, $R_1 = \frac{\hefour\eepn}{\hefour\eep}$,
  $R_2=\frac{\hefour\eepp}{\hefour\eep}$) we assumed factorization and used
  the equations A.1-A.3 listed below:

\begin{align}
\frac{\#pp}{\#np} = \frac{ T_L \cdot R - P_{\textrm{SCX}} \cdot
  \frac{\sigma_{en}}{\sigma_{ep}} }{2 \cdot T_L - 2\cdot P_{\textrm{SCX}} \cdot
  \frac{\sigma_{en}}{\sigma_{ep}} \cdot R }\\ 
\frac{\#pp}{\#p} = \frac{R_1 \cdot \frac{\sigma_{en} }{\sigma_{ep} }
  \cdot \frac{P_{\textrm{SCX}}}{T_L} \cdot T_R - R_2 \cdot T_R }{2\cdot
  (\frac{\sigma_{en}}{\sigma_{ep}} \cdot \frac{P_{\textrm{SCX}}}{T_L} \cdot T_R )^2 -2
  \cdot T^2_R}\\
\frac{\#np}{\#p} = \frac{R_2 - 2 \cdot \frac{\#pp}{\#p} \cdot T_R }{ \frac{\sigma_{en}}{\sigma_{ep}}\cdot \frac{P_{\textrm{SCX}}}{T_L}\cdot T_R }\
\end{align}

\noindent where $\sigma_{ep}$ ($\sigma_{en}$) is the cross section for
electron scattering off the proton (neutron) \cite{paper21}.

%%%%

\end{document}